# Strongly dispersive dielectric properties of high-ScN-fraction ScAlN deposited by molecular beam epitaxy

Vikrant J. Gokhale[1], James G. Champlain[1], Matthew T. Hardy[1], James L. Hart[2], Andrew C. Lang[1], Saikat Mukhopadhyay[1], Jason A. Roussos[3], Shawn C. Mack[1], Gabriel Giribaldi[4], Luca Colombo[4], Matteo Rinaldi[4], Brian P. Downey[1]

1. US Naval Research Laboratory, Washington DC
2. Nova Research Inc, 1900 Elkin St # 230, Alexandria, VA 22308, USA
3. Leidos, 1750 Presidents St., Reston VA 20190
4. Institute for NanoSystems Innovation (NanoSI), Northeastern University, Boston MA 02115

## I. Abstract

We present a comprehensive study of dielectric properties including permittivity, loss, and leakage of high-ScN-fraction ScAlN thin films grown using molecular beam epitaxy (MBE). Dielectric spectroscopy is carried out on high-ScN-fraction (30%-40% ScN fraction) samples from 20 Hz to 10 GHz. We find that real permittivity $\varepsilon'$ increases significantly with increasing ScN fraction; a trend confirmed by density functional theory. Further, $\varepsilon'$ is strongly dispersive with frequency and increasing ScN fraction, with values for $Sc_{0.4}Al_{0.6}N$ varying from 150 down to 60 with increasing frequency. Loss, dispersion, and DC leakage current correspondingly increase with ScN fraction. The high $\varepsilon'$ and strongly dispersive behavior in MBE ScAlN are not observed in a sputter-deposited ScAlN control with a similar ScN fraction, highlighting fundamental differences between films produced by the two deposition methods. Microscopy and spectroscopy analyses are carried out on MBE- and sputter-deposited samples to compare microstructure, alloy, and dopant concentration.

## II. Introduction

The initial development of the ternary alloy scandium aluminum nitride $Sc_xAl_{1-x}N$ (commonly ScAlN, or AlScN) was strongly motivated by its use in radio-frequency (RF) acoustic signal processing systems, since with increasing ScN fraction ($x \to 0.43$) the piezoelectric response of wurtzite ScAlN is up to five times that of AlN [1-6]. ScAlN-based microelectromechanical systems (MEMS) have been used successfully as ultrasonic transducers [7], RF surface and bulk acoustic wave (SAW, BAW) resonators and filters [8-10], and sensors [11]. Ferroelectricity was recently demonstrated in ScAlN [12], leading to strong interest in ScAlN for switching and memory applications [13-18]. Strong optical non-linearities in ScAlN are being investigated, with a goal of developing photonic devices and circuits [19-21]. For a majority of these piezoelectric, ferroelectric, and photonic applications, the deposition method of choice has been sputtering [4, 12, 22]. For MEMS-driven development, the primary goal was to achieve *c*-axis oriented polycrystalline films with the highest stable ScN fraction and the strongest piezoelectric performance, while retaining film texture, low roughness, low density of abnormally oriented grains (AOGs), low stress, and compatibility with MEMS process flows. Recently, process development of molecular beam epitaxy (MBE) for the growth of single-crystal ScAlN on GaN, SiC, Si, or epitaxial transition metal nitride (TMN) metallic surfaces has enabled lattice-matched ScAlN/GaN high electron mobility transistors (HEMTs) with promising current density, output power density, and RF characteristics [23-26]. MBE ScAlN can also be used for MEMS and ferroelectric applications: SAW/BAW resonators operating at frequencies up to 110 GHz have been reported [27-32], and ferroelectric behavior has been demonstrated [25]. Depending on the technique used, a variety of growth/deposition environments and growth kinetics are expected, likely resulting in very different microstructure and properties for the ScAlN film [26, 33-37]. Quantifying and understanding the causal or correlative technique-structure-property relationships will be critical for material development and future applications.

Here, we focus on characterizing the dielectric properties of ScAlN and their dependence on deposition technique, ScN fraction, and film microstructure. Each application type noted above has other important parameters (piezoelectric coefficients, band offsets, coercive fields for ferroelectric switching, second-order susceptibility, etc.), but a complete understanding of the dielectric permittivity and loss in ScAlN films is useful for all aforementioned applications. To date, several experiments have directly measured or extracted the dielectric constant or the real permittivity ($k$ or $\varepsilon'$) of ScAlN at single or a few fixed frequencies [3, 15, 38, 39], yet there is a lack of comprehensive broad-spectrum experiments and analyses. A notable exception is a recent study by Casamento *et al.* [37], which tested MBE-grown ScAlN for ScN fractions ranging from 17% – 25%, and across a frequency range from $10^3$ Hz – $10^7$ Hz. That study demonstrates the potential for ScAlN as a high-$k$ gate dielectric material, concluding that $Sc_{0.25}Al_{0.75}N$ has a relatively high real permittivity of $\varepsilon' \sim 22$, roughly twice that of GaN or AlN. Interestingly, Casamento *et al.* also found that while real permittivity was relatively low and constant across frequencies for $Sc_{0.17}Al_{0.83}N$ and $Sc_{0.2}Al_{0.8}N$ ($\varepsilon'(\omega) \sim 15$ and $\varepsilon'(\omega) \sim 16$ respectively), it was higher ($\varepsilon'(\omega) \rightarrow 22$) and weakly dispersive as a function of frequency for $Sc_{0.25}Al_{0.75}N$. The $Sc_{0.25}Al_{0.75}N$ film also has higher dielectric loss and leakage current. These results warrant further investigation into the permittivity of MBE ScAlN, especially for $x > 0.25$.

In this report, we use dielectric spectroscopy to measure the dielectric properties of high-ScN-fraction ($x \sim 0.3 - 0.4$) $Sc_xAl_{1-x}N$ thin films grown by MBE. The films are characterized over a wide range of frequencies (almost 9 decades, from 20 Hz to 10 GHz). We characterize the complex dispersive permittivity of epitaxial ScAlN, the dielectric loss, and current leakage through the films, all as a function of ScN fraction and frequency. We compare the measured dielectric data with structural characterization of the films using various analytical techniques and use *ab initio* density functional theory (DFT) to provide a theoretical validation of the measured trends. We compare findings on the MBE films with a high quality sputtered ScAlN thin film deposited using an optimized RF-magnetron sputtering process.

## III. Dielectric Properties of thin film ScAlN

Dielectric spectroscopy, or impedance spectroscopy, is a useful technique for evaluating the complex permittivity of dielectric materials. In its simplest frequency-domain implementation, this technique subjects a parallel plate capacitor made of the dielectric material under investigation to a small-signal stimulus over a wide range of frequencies given by $\omega = 2\pi f$. The measured impedance ($Z(\omega)$) or admittance ($Y(\omega)$) is modeled as a parallel combination of a capacitor and a conductor, with $1/Z(\omega) = Y(\omega) = G + j\omega C(\omega)$, where $G$ and $C$ are the dissipative conductance and the non-dissipative capacitance. Alternately, the model can be described as a complex capacitance written as $C(\omega) = A\varepsilon_0\varepsilon(\omega)/t$, where $A$ and $t$ are the area and thickness of the capacitor, $\varepsilon_0$ is the vacuum permittivity, and $\varepsilon(\omega) = \varepsilon'(\omega) - j\varepsilon''(\omega)$ is the frequency dependent complex permittivity of the dielectric. Dielectric spectroscopy of a capacitor represents a procedurally simple but powerful characterization technique for novel dielectric materials, and we use it here for characterizing MBE ScAlN. Three MBE ScAlN samples (E1, E2, E3) and one sputter-deposited control sample (S1) are examined (Table I). Figure 1 (a) and (b) show a cross-sectional schematic and layout of the 'metal-insulator-metal capacitor' (MIMCAP) heterostructure respectively. The top electrodes are in a concentric 'large area – small area' (LASA) configuration. The LASA-MIMCAP configuration allows us to acquire dielectric measurements of a thin film without needing to pattern and selectively etch the dielectric film and contact the bottom electrode directly; high-quality MBE ScAlN can be a difficult material to wet-etch selectively and controllably [30, 40]. The LASA test structures require only one mask and process step (defining the top electrodes using optical lithography followed by electron beam evaporation of Ni/Au and liftoff) and can be included in most process flows. Further, unlike other extraction techniques [41], LASA-MIMCAP data do not need differential measurement and all analysis can be carried out on a single device. Figure 1 (b) shows a top view of the LASA configuration with circular electrodes of inner radii $r_1$ ranging from 15 μm to 85 μm. We perform dielectric spectroscopy across 9 decades of frequency using an impedance analyzer (ZA) and vector network analyzer (VNA). Details of ScAlN film deposition, MIMCAP fabrication, and measurement procedure are provided in the Methods section.

The lumped impedance model of the MIMCAP $Z_T$ is shown in Figure 1 (c). Given the physical configuration of the MIMCAP heterostructures, we can separate the total impedance into the impedances of the inner capacitor and the outer capacitor, along with the impedance of the annular gap, and any contact resistance associated with the probes (Figure 1 (d)). The measured impedance is normalized with respect to area and frequency for comparison across devices/samples and is rewritten as $(Z_T^{-1}/\omega A) = (G/\omega A) + j(C/A)$. Figure 1 (e) and (f) show normalized capacitance $(C/A)$ and conductance $(G/\omega A)$ respectively for all samples, for MIMCAPs with $r_1 = 50$ μm (Data for all MIMCAPs are shown in Supplementary Information). It is immediately apparent that sample S1 has a nearly constant capacitance across the entire range and a low conductance while the MBE samples E1 and E2 (both nominally 32% ScN) and E3 (nominally 40% ScN) have higher capacitance. Capacitance for all MBE samples is dispersive and decreases with increasing frequency instead of maintaining a constant flat value. The trends for conductance are similar: S1 has the lowest conductance while MBE samples have higher levels of conductance, with sample E3 demonstrating the highest values.

A full mathematical derivation and analysis of the LASA-MIMCAP model will be published elsewhere [42]; a short summary is presented in the Methods. We use this model to extract complex permittivity of ScAlN. Extracted real permittivity $\varepsilon'(\omega)$ of the ScAlN layer for all samples is shown in Figure 2 (a) for MIMCAPs with $r_1 = 50$ μm (All data are provided in Supplementary Information). In addition, we plot $\varepsilon'(\omega)$ data from the study by Casamento *et al* [37], with MBE-grown samples 17%, 20%, and 25% ScN-fraction ScAlN samples (labeled C1, C2, and C3 respectively). The permittivity of MBE ScAlN films increases with increasing ScN fraction. In contrast, the sputtered sample S1 has a flat real permittivity $\varepsilon'(\omega) \sim 17$ across the entire range of frequencies. To quantify the degree of dispersion in the data, we fit data to a power law function $\varepsilon'(\omega) = \alpha \omega^{-\beta}$, commonly used when describing dielectric functions [43]. The fits are shown by lines in Figure 2 (a), and the corresponding fit coefficients $\alpha$ and exponents $\beta$ are shown in Figure 2 (b) and (c) respectively as a function of ScN fraction. Sample S1 has a negligible dispersion, with $\varepsilon'(\omega) \sim 16.7$ across the entire frequency range. The clear trends for $\alpha$ and $\beta$ as a function of ScN function demonstrates that $\varepsilon'$ both increases and gets more dispersive for MBE ScAlN. The dispersion seen in the MBE films could be eventually attributed to one or a combination of several possible dielectric relaxation processes; however, at this point we do not attempt to separate these processes. The $\varepsilon'(\omega) = \alpha \omega^{-\beta}$ power law fits here are purely empirical; future experiments shall aim to separate/isolate the root cause behind the dispersion. We draw no quantitative conclusions from the consistency between MBE samples from this study and those from Casamento *et al* [37], grown in a different lab, on different substrates. Qualitatively, we note that there is a consistent trend for $\varepsilon'(\omega)$ across MBE ScAlN from two independent, non-overlapping studies, and that large differences, unexplored to date, exist between the dielectric behavior of MBE and sputtered ScAlN.

To verify the observed high permittivity in the MBE ScAlN, we used *ab initio* density functional theory (DFT) (Methods). Above ~40% ScN fraction, ScAlN is expected to transition from a hexagonal to cubic phase, with an accompanying and drastic shift in many material properties[1-3, 44, 45]. In our DFT calculations for wurtzite $Sc_{0.5}Al_{0.5}N$, this phase transition manifests as imaginary modes in the phonon band structure for ScN fractions greater than 43%, validating the model (Figure 3(b)). The DFT-calculated real permittivity is shown in Figure 3(a) along with selected data from this experiment and from the literature [2, 3, 37, 38]. The measured values of real permittivity for the MBE films converge to the DFT calculated values. Our DFT calculations predict higher values of permittivity than the study by Caro *et al* [5] but agree well with converged values of the MBE ScAlN data. Permittivity for S1 and several other sputtered samples from the literature are closer to the Caro model. This is an indicator that different DFT approaches are suited for different types of film microstructure and will be a topic for deeper investigation.

The extracted imaginary permittivity $\varepsilon''(\omega)$ for all samples is shown in Figure 4(a). As with $\varepsilon'(\omega)$, the sputtered sample S1 has the lowest $\varepsilon''(\omega)$, with increasingly higher and more dispersive values for the MBE samples as the ScN fraction increases. In the low frequency dispersion (LFD) region below ~1 kHz, we can see higher apparent values of $\varepsilon''$, especially stark for S1. This apparent $\varepsilon''(\omega)$ is caused by DC leakage current and can be modeled as $\varepsilon''_{LFD}(\omega) \propto \sigma_{DC} \varepsilon_0^{-1} \omega^{-1}$ where $\sigma_{DC}$ is the DC conductivity [43] (measured DC leakage for all samples is quantified below). At higher frequencies, the $\varepsilon''(\omega)$ values for S1 are relatively flat. Even though they are leakier (see below) the MBE films do not show a clear $\omega^{-1}$ trend in the LFD, which indicates that $\varepsilon''(\omega)$ is likely dominated by other dielectric relaxation processes with sub-linear $\omega^{-\beta}$ trends where $\beta \ll 1$. We do not yet attempt to fit the measured MBE ScAlN data since there are several possible LFD mechanisms with sub-linear $\omega^{-\beta}$ trends that could cause losses in the LFD region [43]. Future experiments will attempt to separate the loss mechanisms. The dielectric loss tangent for the ScAlN films, defined as $\tan\delta = \varepsilon''(\omega)/\varepsilon'(\omega)$ is shown in Figure 4(b). It follows from the permittivity data above that the loss tangent of the MBE films increases with increasing ScN content. The loss tangent of MBE film C3, calculated from [37], follows the trends seen in our MBE samples. In contrast, the loss tangent for sputtered film S1 is as low as 2×10$^{-3}$ for frequencies > 1 kHz. Data >1 GHz can be limited by the geometrically defined MIMCAP cutoff frequencies; the apparent sharp roll-off in $\varepsilon'(\omega)$ or increases in $\varepsilon''(\omega)$ and $\tan\delta$ at higher frequencies are not intrinsic to the material (Supplement). The average DC current density $J_{DC}$ and DC conductivity $\sigma_{DC}$ for all MIMCAPs on all samples is shown in Figure 4(c)-(d). Here, $\sigma_{DC}$ represents the conductivity for the entire dielectric stack including the nucleation layers where applicable. The trends in $J_{DC}$ and $\sigma_{DC}$ with respect to ScN fraction are consistent with the trends in $\varepsilon'$ and $\varepsilon''$. S1 has negligible leakage compared to the MBE samples, and the MBE films demonstrate increasing DC leakage with increasing ScN fraction. While there is an expected narrowing of the bandgap with increased ScN fraction [44], the large difference in leakage between S1 sputtered $Sc_{0.3}Al_{0.7}N$ (S1) and MBE $Sc_{0.32}Al_{0.68}N$ (E1, E2) indicates that other factors related to microstructure or impurity doping could be dominant in the MBE films. All MBE samples were grown using a 'high purity' Sc source (Methods); lower purity Sc sources have been associated with higher levels of DC leakage [36, 46]. Two key observations can be made considering the $\varepsilon'(\omega)$, $\varepsilon''(\omega)$, and $\sigma_{DC}$ data together. First, while $\sigma_{DC}$ might account for $\omega^{-1}$ behavior in the LFD region, it does not explain the sublinear $\omega^{-\beta}$ dispersion in $\varepsilon'(\omega)$ or $\varepsilon''(\omega)$ at higher frequencies up to ~1 GHz. This high frequency dispersion must originate from other processes caused either by imperfections (impurities, vacancies, deep traps, hopping conduction, etc.) in the MBE films, or by dipolar relaxation processes inherent to the microstructure and texture of the MBE ScAlN. Second, the significantly higher magnitude of $\varepsilon'$ with increasing ScN fraction cannot be explained solely by $\sigma_{DC}$ and must originate from other intrinsic dielectric processes in the MBE ScAlN.

## IV. Film microstructure and impurity concentration

In order to correlate deposition technique and structural/compositional properties with the measured dielectric properties, we characterized the ScAlN films via secondary ion mass spectroscopy (SIMS), X-ray diffraction (XRD), atomic force microscopy (AFM), and transmission electron microscopy (TEM). SIMS was performed on samples E2, E3 and S1 to measure the alloy composition of ScAlN, along with any impurities. Figure 5(a-c) shows the Sc, Al, and Nb profiles of the three imaged samples, with composition values and film thicknesses close to nominal values. These values are confirmed by energy-dispersive X-ray spectroscopy (EDS) (Table I)) performed during TEM imaging. Note that error bounds for EDS and SIMS are ± 5%. The oxygen and fluorine profiles in Figure 5 (d-e) show that MBE films have ~ 10× lower oxygen concentration and between 5×-10× lower fluorine concentration as compared to the sputtered film. Neither the presence or relative concentration of impurities explains the higher leakage and dispersive behavior observed in MBE ScAlN films as compared to sputtered ScAlN. The SIMS impurity analysis also clarifies another important point: impurity concentration profiles for E2 (32% ScN) and E3 (40% ScN) are nearly identical even though nominal and measured ScN profiles are markedly different, unlike prior reports [46]. This effectively eliminates Sc source contamination as a likely cause for leakage, loss, or dispersion in this sample set.

The XRD full width at half maximum (FWHM) values of the 0002 ScAlN peak for all samples is included in Table I. The FWHM values for both MBE and sputtered films deposited on epitaxial NbN are comparable to or much better than typically reported sputtered ScAlN films in the literature. Sample E3 has the broadest (worst) FWHM, as expected from previous reports [1, 34]. AFM scans on the wafers show low surface roughness and no correlation with ScN fraction. We perform TEM imaging and selected area electron diffraction (SAED) on samples E2, E3, and S1. Table 1 provides the in-plane and out-off-plane lattice parameters (*a* and *c* respectively) determined from SAED measurements. The data and the *c/a* ratios match expected trends as a function of ScN fraction [47, 48]. Figure 6(a-c) shows representative bright field TEM (BF-TEM) images of the samples. All three ScAlN films show significant intensity fluctuations, which indicates the presence of defects and disorder. For the two MBE films, the defect density is highest near the ScAlN/AlN interface and decreases near the top surface. Qualitatively, E3 has a higher defect density compared to E2, indicating that higher ScN content is correlated with increased disorder and defects. In these BF-TEM images, the observed defects (i.e. the intensity variation) is mainly attributed to microcrystal rotations and low-angle grain boundaries (mosaicity) [49]. This is evidenced by atomic-resolution high angle annular dark field scanning TEM (HAADF-STEM). An example HAADF-STEM image of S1 is provided in Figure 6(d). The structure varies on a length scale of several nanometers, oscillating between regions where atomic columns are clearly resolved, and regions where the atomic structure is blurred out. These images suggest that the ScAlN films are composed of many nanoscale crystals. In regions where the atomic columns are well-defined, the crystal is well-aligned with the STEM optic axis. In regions where the atomic structure appears blurred, the crystal is slightly mis-aligned with the optic axis, or there are overlapping microcrystals which are laterally or rotationally offset.

To better compare the crystal quality, we compare atomic-resolution HAADF-STEM images and SAED data. STEM images for E2, E3, and S1 are shown in Figure 6(e-g), respectively, with each image taken ~100 nm away from the bottom interface. There are clear differences in the crystal quality and defect density. The lower ScN fraction MBE film (E2) shows defect-free regions on the order of ~10 nm. The higher ScN fraction MBE film (E3) has a much higher defect density, while S1 has a defect density intermediate between E2 and E3. The E2 sample shows the sharpest diffraction spot, indicating the highest crystal quality. The E3 film clearly has the broadest diffraction spot (lowest crystal quality), while the sputtered S1 film is again intermediate between E2 and E3. Based on HAADF-STEM analysis, the ScAlN/AlN interface for E3 is comparable to that of E2 while S1 shows a sharper ScAlN/NbN interface (Supplementary). STEM and SAED were also used to assess the presence of secondary phases in the ScAlN films, specifically, cubic ScAlN. Figure 6(h-j) show SAED patterns of E2, E3, and S1, respectively. For E2, the SAED shows no cubic reflections, nor are any cubic grains observed with STEM imaging. In contrast, both E3 and S1 show cubic reflections in their SAED patterns [50]. For both films, we observe very broad cubic reflections, suggesting many nanoscale cubic inclusions within the wurtzite film (white arrows in Figure 6(i-j). In addition to these broad reflections, for the sputtered film S1, we also observe one set of very sharp cubic reflections, indicating the presence of a much larger cubic grain. For film E3, we mapped the cubic inclusions in real-space with STEM imaging. Figure 6(k) shows a large field of view (~60 nm) atomic-resolution STEM image of E3. The lower right inset provides the image Fourier transform (FT). In addition to the wurtzite [11-20] spots, the FT shows several additional reflections, all of which are consistent with cubic ScAlN. We mask these spots and perform an inverse FT, yielding the cubic phase map shown in Figure 6l. The cubic grains are several nm in diameter, and appear randomly dispersed throughout the film, consistent with the SAED data. Note that the intensity in Figure 6(l) has been enhanced to highlight the cubic grains, and this image should not be used to assess the volume fraction of cubic grains in E3. Based on SAED intensity analysis, the volume fraction of cubic phases appears low, on the order of a few percent. While offset or rotated grains, interfacial roughness, or cubic inclusions could explain dielectric loss or leakage in ScAlN films, we find that these effects are not obviously correlated with the measured dielectric data above.

## V. Conclusion

This work presents a comprehensive and systematic study of the dielectric properties of high-ScN-fraction MBE-deposited ScAlN films. We find significantly higher values of real permittivity than previously reported (~100 at 1 MHz for MBE $Sc_{0.4}Al_{0.6}N$), making it attractive as a high-$k$ dielectric material that can be epitaxially integrated with GaN HEMT electronics. The permittivity was highly dispersive with frequency, with the dispersion rate increasing with ScN fraction. Such strongly dispersive trends in permittivity and their dependence on ScN fraction are reported for the first time. We correlate dielectric permittivity with analyses of the microstructure, film quality, and impurity profile. While higher defect density or higher impurity doping might provide a facile hypothesis for the MBE ScAlN films being more lossy and dispersive than the sputtered ScAlN control, we do not find conclusive evidence to support such explanations. On the contrary, we find that that film quality of the sputtered and MBE films is similar and impurity concentration levels are higher in the sputtered film. While the leakage influences the low frequency dispersion of permittivity and loss, it does not satisfactorily explain either the high values or the strong dispersion in permittivity across a wide range of frequencies. Comparisons with DFT calculations show that the permittivity trends for MBE ScAlN qualitatively agree with our measurements; previous models tend to underestimate the permittivity. We show that the dielectric properties of MBE ScAlN are substantively different from those of sputtered ScAlN, even when the ScN fraction is similar. As the growth process is further optimized for dielectric properties, we expect to gain better insight into these differences and finer control over the properties themselves. The large application space for high-ScN-fraction, MBE-grown thin-film ScAlN makes it imperative that we understand and study these properties.

# VI. METHODS

### a. Growth of ScAlN thin films

In prior work, we have developed processes to grow high ScN fraction ($x \geq 18$) $\text{Sc}_x\text{Al}_{1-x}\text{N}$ films and metallic/superconducting transition metal nitrides (TMNs) using plasma assisted molecular beam epitaxy (MBE) [24, 26, 33, 34, 40, 51-54]. All MBE samples used in this study (E1, E2, and E3 in Table I) were grown in an Omicron PRO-75 MBE system with an RF plasma source for active nitrogen (N*). The chamber is fitted with an electron beam evaporator to produce Nb flux and effusion cells for Al and Sc. The Sc source is nominally 99.999% pure using trace rare earth metals basis, and 99.97% pure overall, per the manufacturer. We use 100 mm, double-side polished semi-insulating 4H-SiC substrates and grow a NbN TMN layer as the bottom electrode (nominally 50 nm thick) at nominally 850 °C, and a growth rate of 0.26 Å/s [53]. Next, we grow a two-step AlN nucleation layer with a total nominal thickness of 30 nm. The first step is grown N-rich with a III/V ratio of ~0.5 at a nominal substrate temperature of 770 °C. However, due to the change in emissivity of the NbN layer, the surface temperature at the start of the growth is estimated to be 950 °C [54]. The Al cell temperature is then ramped up and the substrate temperature decreased while growing the second half of the AlN NL metal rich (III/V = 1.1), giving a final estimated substrate surface temperature of 750 °C and growth rate of 0.7 A/s. Growth is then interrupted and the substrate temperature is ramped down to 420 °C for the ScAlN growth using a III/V ratio of 0.93 and a growth rate of 0.64 A/s [33]. The composition for each sample is adjusted by changing both the Al and Sc flux to maintain a constant growth rate and III/V ratio  For the $\text{Sc}_{0.40}\text{Al}_{0.60}\text{N}$ sample (E3), the first 25 nm of the ScAlN layer is grown using a linear composition grade from $\text{Sc}_{0.32}\text{Al}_{0.68}\text{N}$ to $\text{Sc}_{0.40}\text{Al}_{0.60}\text{N}$ to suppress formation of rock-salt phases during ScAlN nucleation and reduce the tensile stress [34, 50]. Sputtered sample S1 starts with the same epitaxial NbN/SiC template as all the MBE samples. An Evatec © Clusterline 200 tool with a 12" $\text{Sc}_{0.30}\text{Al}_{0.70}\text{N}$ casted target were used to deposit 290 nm of ScAlN using optimized conditions described in prior work [55].

### b. Impedance and leakage measurements

Low frequency impedance measurements from 20 Hz to 5 MHz are acquired using a Keysight E4990A impedance analyzer (ZA) and needle probes for contacting the top MIMCAP electrodes. High frequency impedance measurements from 300 kHz to 10 GHz are acquired using a Keysight E5071C vector network analyzer (VNA) and RF probes in a ground-signal-ground (GSG) configuration. Data in the overlap region between ~ 300 kHz to ~5 MHz are noisy near the instrument cutoff frequencies. No attempt is made to adjust or normalize measured impedance or other derived quantities across the two instruments. No data are discarded or modified but the subsequent solution for $\varepsilon(\omega)$ is not well conditioned in these regions. As-measured, raw impedance data across the entire spectrum are included in the Supplementary Information. DC leakage and current density measurements are acquired with a Keithley SCS4200 parameter analyzer.

### c. LASA-MIMCAP design and complex permittivity extraction

For LASA-MIMCAPs used in this work, the inner electrode is a small circle of radius $r_1$ ranging from 15 µm to 85 µm. The radius of the outer annulus is set to $r_2 = 1.4 \times r_1$, although a fixed ratio is not necessary. The annular gap $g = (r_2 - r_1)$ between the electrodes is much larger (20× – 100×) than the total thickness of the dielectric heterostructure $(g >> t)$; allowing us to neglect lateral capacitance. The bottom NbN electrode is capacitively coupled to the large area outer electrode on the top surface, with the area of the large outer electrode at least ~100× larger than the area of the inner electrodes. The LASA-MIMCAP can be modeled as the sum of the impedances of the small (inner) and a large (outer) parallel plate capacitor in series, connected by the impedance of the conductive bottom plate, with the additional series resistance $R_s$ accounting for the resistance of the top electrodes, any cable resistance, and the probe contact resistance. The total frequency dependent complex impedance $Z_T(\omega)$ of the MIMCAPs can be written as the sum of its constituent impedances. For the circular and annular geometries involved, the exact analytic expression for the total impedance is given by:

$$Z_T(\omega) = Z_{inner}(\omega) + Z_{gap}(\omega) + Z_{outer}(\omega) + R_s$$

$$Z_T(\omega) = \frac{1}{j\omega C_T \pi r_1^2} \frac{\alpha_1 J_0(\alpha_1)}{2 J_1(\alpha_1)} + \frac{1}{2\pi \sigma_{SH}} \ln\left(\frac{r_2}{r_1}\right) + \frac{1}{2\pi \sigma_{SH}} \frac{1}{\alpha_2} \frac{H_0^{(1)}(\alpha_2)}{H_1^{(1)}(\alpha_2)} + R_s$$

Here, $C_T = \varepsilon_0 \varepsilon / t$ is the total capacitance per unit area of the dielectric heterostructure between the top and bottom conductors, and thus the measured impedance can be modeled as $Z_T(\varepsilon(\omega))$. Here, $\varepsilon_0$ is the vacuum permittivity, $t$ is the thickness of the dielectric heterostructure, $r_1$ and $r_2$ are the radii of the inner and outer contacts, respectively; $\alpha_n = \sqrt{-j\omega \tau_n}$ is a unitless factor with $\tau_n = C_T r_n^2 / \sigma_{SH}$ defining a characteristic time constant associated with the inner ($n = 1$) or outer ($n = 2$) contact. $\sigma_{SH}$ and $R_{SH}$ are the lateral sheet conductance and sheet resistance of the bottom conductor ($\sigma_{SH} = R_{SH}^{-1}$); and $R_s$ is a small lumped series resistance accounting for any resistance outside the MIMCAP (e.g., resistance of thick metallic top contacts, probe contact resistance and cable resistance). The functions $J_n(\alpha_1)$ and $H_n^{(1)}(\alpha_2) = J_n(\alpha_2) + jY_n(\alpha_2)$ are Bessel functions of the first kind and Hankel functions of the second kind, respectively, while $Y_n(\alpha_2)$ are Bessel functions of the second kind. Note that $Z_{inner}$ and $Z_{outer}$ include the resistance of the bottom NbN conductor in the inner cylinder and outer toroid respectively. A low frequency approximation of the MIMCAP model (i.e., $\omega\tau_1 < \omega\tau_2 \ll 1$) is given by:

$$Z_T = \left[\frac{1}{8\pi\sigma_{SH}} - j\frac{1}{\omega C_T \pi r_1^2}\right] + \left[\frac{1}{2\pi\sigma_{SH}} \ln\left(\frac{r_2}{r_1}\right)\right] + \left[\frac{1}{2\pi\sigma_{SH}}\left\{\frac{1}{2}\ln\left(\frac{4}{\omega\tau_2}\right) - \gamma_e\right\} - j\frac{1}{8\sigma_{SH}}\right] + R_s$$

Where $\gamma_e$ is Euler's constant. For the smallest radius MIMCAPs in this work, the low frequency approximation is valid up to ~10 GHz, with larger devices exhibiting lower cutoff frequencies as $\omega\tau_1, \omega\tau_2 \to 1$ (see Supplementary Information). A similar configuration used in the literature does not completely model $Z_{outer}$, requiring differential measurements across devices, introducing additional sources of error without significant advantage [41, 56]. In contrast, our approach presents an exact solution for $Z_T(\varepsilon(\omega))$ using a single measurement. We solve for the complex permittivity of the dielectric heterostructure $\varepsilon(\omega)$ by using the symbolic toolbox in MATLAB, and using measured complex impedance data, directly measured parameters such as film thickness and sheet resistance from Table I, and the LASA-MIMCAP model above. The only arbitrary free parameter is the contact resistance $R_s$, which assumes values between 1Ω and 6Ω, and does not significantly change the extracted values (see Supplementary Information). Note that both the exact form and the low-frequency approximation of $Z_T(\varepsilon(\omega))$ yield the same solutions; the exact form can take longer to converge on the noisier fringes of the measurement range. We isolate the permittivity of the ScAlN layer by assuming that the AlN nucleation layer (when present in the MBE samples) has a constant, real, and non-dispersive permittivity ($\varepsilon_{AlN}(\omega) = 10 - j0$). This simplification assumes that any dispersion and losses observed in the measurements are concentrated in the ScAlN layer, and not arising from either the AlN layer (if present) or any of the interfaces. This approach gives us the worst-case scenario for the loss in the ScAlN layer.

d. Density function theory calculations

To generate the ScAlN structures with varying ScN fraction, we used MedeA [57] to randomly substitute Al atoms by Sc atoms in a 2×2×2 supercell with 32 atoms. This approach is similar to experimentally validated DFT studies on other substitution doped solid state materials [58]. However, it is in marked contrast with previous ScAlN studies which used special quasi random structures (SQS) to generate the ScAlN [5, 38, 59]. In our DFT approach, the lattice and internal coordinates of wurtzite ScAlN were optimized with the projector augmented wave (PAW) method [60] within the generalized gradient approximation (GGA) [61, 62] as implemented in the Vienna Ab initio Simulation Package (VASP) [63]. An energy cutoff of 520 eV and a convergence criterion of $10^{-6}$ eV for energy and $10^{-4}$ eV/Å for energy gradient was used. In order to eliminate any error arising from the size of the supercell used in the calculation, we repeated the calculation with a 3×3×3 supercell (108 atoms) with similar results.

e. Transmission electron microscopy

TEM samples were prepared with a Thermo Fisher Scientific Helios G3 focused ion beam with final milling performed at 5 kV with Ga ions. Specimens were further cleaned in a Fischione nanomill with Ar ions down to 1 kV. The BF-TEM, SAED, and EDS data was acquired on a Thermo Fisher Scientific Talos operated at 200 kV. To extract the ScAlN lattice parameters from the SAED data, we collect patterns which include both the ScAlN film and the underlying SiC substrate. We calibrate the pattern based on the substrate peaks, allowing accurate measurement of the ScAlN lattice parameters.  In BF-TEM image analysis defect free crystals with uniform strain exhibit uniform intensity. Defects and non-uniform strains lead to variation of BF-TEM image intensity. SAED measurements are used to probe the film quality averaged across a ~500 nm diameter aperture.  The width of the $(1\bar{1}0\bar{3})$ reflection is used as a proxy for crystalline quality and mosaicity, with a sharper spot indicting higher crystal quality. The atomic-resolution STEM imaging was performed on a Nion UltraSTEM 200X, operated at 200 kV with a 30mrad convergence angle.

## VII. Acknowledgements

We thank Dr. Jeff LaRoche, Dr. Eduardo M. Chumbes, and Dr. Adam Peczalski, (RTX) for their support on this project and for many invaluable discussions, and Dr. Jeffrey P. Calame (NRL) for his knowledge and insights into dielectric materials and behavior. This work was funded by the Defense Advanced Research Projects Agency, Microsystem Technology Office (DARPA MTO) under the COmpact Front-end Filters at the ElEment-level (COFFEE) program, and the NRL Base Program. We are grateful to Dr. Benjamin Griffin, Dr. Todd Bauer, and Dr. Zachary Fishman of DARPA for their programmatic and funding support.

## VIII. Disclaimer

The views, opinions and/or findings expressed are those of the author and should not be interpreted as representing the official views or policies of the Department of Defense or the U.S. Government.

# IX. FIGURES & TABLES

Table I: A summary of the ScAlN films used in this study, grown using molecular beam epitaxy (MBE) or deposited via sputtering. Structural characterization techniques used to characterize samples include X-ray diffraction (XRD), energy-dispersive X-ray spectroscopy (EDS), secondary-ion mass spectrometry (SIMS), transmission electron microscopy (TEM), selected area electron diffraction (SAED), and atomic force microscopy (AFM). Room temperature sheet resistance of the NbN electrodes is measured by contactless mapping.

| Sample ID | Deposition Technique | ScN Fraction* | | | XRD 0002 ScAlN | AFM (5×5 µm²) | Measured Lattice Parameters | | | Film Thickness ‡ | | | Sheet Resistance $R_{SH}$† |
|---|---|---|---|---|---|---|---|---|---|---|---|---|---|
| | | Nominal | EDS | SIMS | FWHM | $r_a$ | a | c | c / a | NbN | AlN | ScAlN | NbN |
| | | % | % | % | ° | nm | Å | Å | - | nm | nm | nm | Ω/□ |
| E1 | MBE | 32 | - | - | 1.28 | - | - | - | - | 52 | 47 | 108 | 15.7 |
| E2 | MBE | 32 | 29 | 29.3 | 0.88 | 0.89 | 3.29 | 4.93 | 1.50 | 48 | 31 | 142 | 37.2 |
| E3 | MBE | 40 | 36 | 35.8 | 2.25 | 1.02 | 3.32 | 4.88 | 1.47 | 50 | 31 | 139 | 30.7 |
| S1 | Sputter | 30 | 28 | 30.3 | 1.38 | - | 3.26 | 4.97 | 1.52 | 48 | N/A | 290 | 14.2 |

\* Error bounds for EDS and SIMS data: ± 5%
‡ Film thickness measurements from TEM; error bounds: ± 1 nm
† Sheet Resistance $R_{SH}$ error bounds ± 1.5 Ω/□. All measurements at room temperature

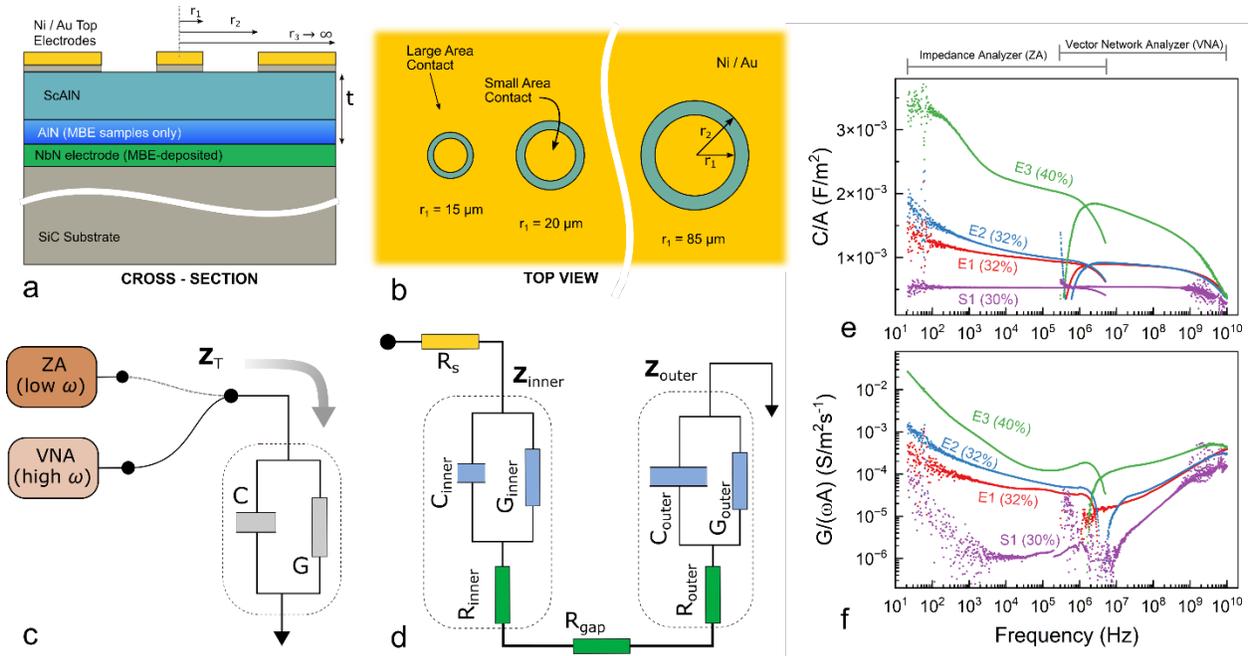

Figure 1: **Test Structures and Measured Impedance.** (a) Cross-sectional view of the ScAlN MIMCAP devices used in this work. (b) Top view of the array of LASA – MIMCAPs used in this work, with input (inner) electrode radius $r_1$ ranging from 15 µm to 85 µm. (c) The total MIMCAP impedance $Z_T$ can be modeled as an ideal capacitor $C$ in parallel with a conductance $G$ which represents the total dielectric losses in the capacitor. We measure $Z_T$ with an impedance analyzer (ZA) at low frequencies and a vector network analyzer (VNA) at high frequencies. (d) The LASA – MIMCAP configuration here can be further expanded into a series combination of the inner and outer capacitors, the resistances of each bottom contact, and the annular resistance of the bottom electrode in the gap. The additional series resistance $R_S$ accounts for the resistance of the Ni/Au top electrodes and any contact / cable resistance. The total impedance can be normalized and written as $(Z_T^{-1}/\omega A) = (G/\omega A) + j(C/A)$. Normalized values of (e) measured capacitance and (f) measured conductance for a MIMCAP of inner radius $r_1 = 50$ µm, across 9 decades of frequency (from 20 Hz to 10 GHz), for all samples. The normalized capacitance $C/A$ exhibits marked dispersion for all MBE samples (E1 – E3), in stark contrast with the near-ideal flat capacitance for the sputtered samples S1. The measured data for $Z_T(\omega)$ on all samples and MIMCAP devices are used to extract the complex permittivity $\varepsilon(\omega) = \varepsilon'(\omega) - j\varepsilon''(\omega)$.

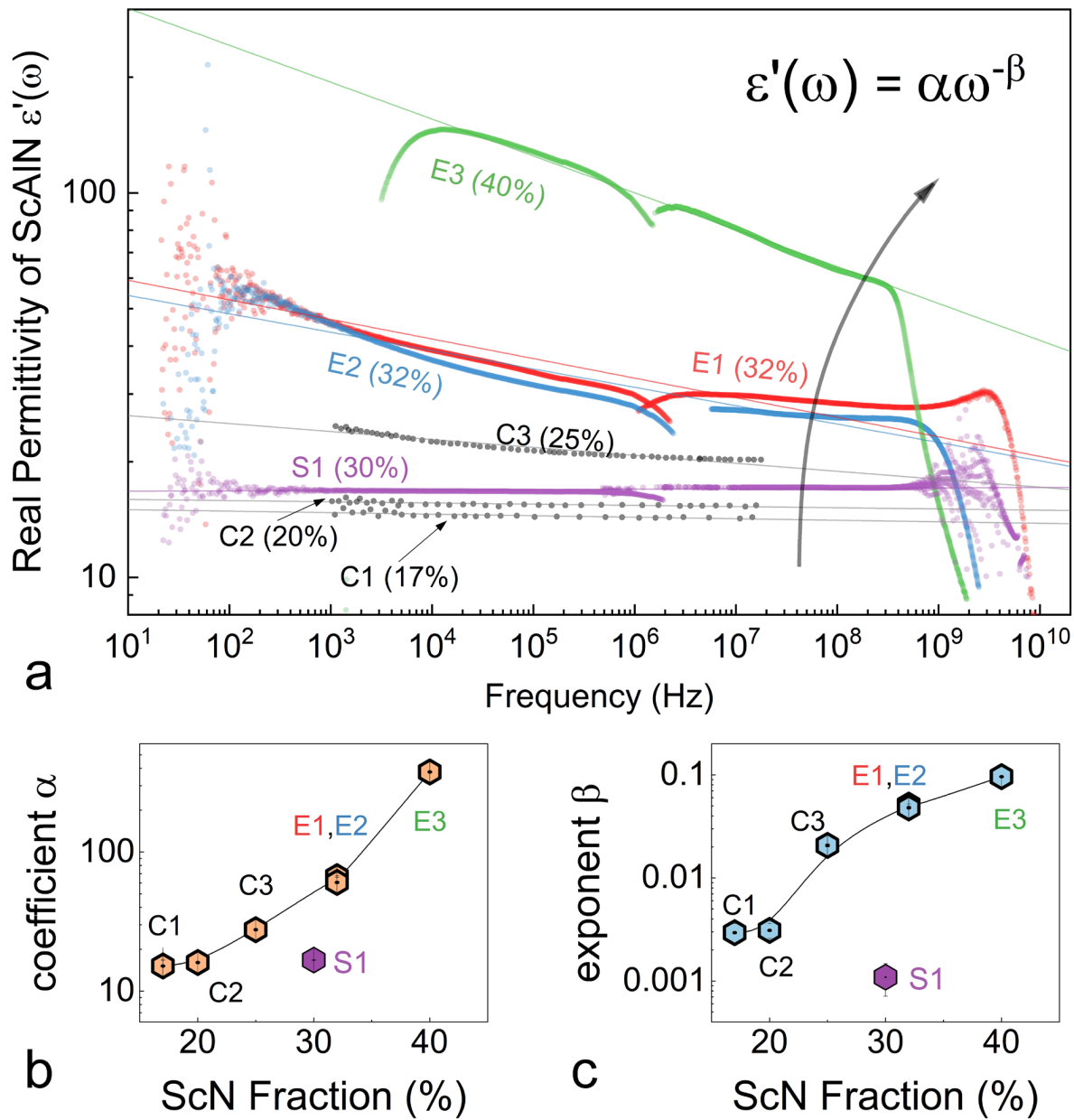

Figure 2: **Dispersive real permittivity of ScAlN as a function of frequency and ScN fraction** (a) Extracted values of real permittivity $\varepsilon'(\omega)$ for MIMCAPs of inner radius $r_1 = 50$ μm, for all samples. In addition to our data, we plot $\varepsilon'(\omega)$ data from the study by Casamento *et al* [37]. Here, C1, C2, and C3 are MBE-grown ScAlN films with ScN fractions of 17%, 20%, and 25% respectively. All MBE films can be fit to power-law trends of the form $\varepsilon'(\omega) = \alpha\omega^{-\beta}$. The magnitude and dispersion trends of the MBE films from [37] along with our data stand in stark contrast with the flat, non-dispersive, lower permittivity of the sputtered ScAlN in this work and in the literature. The (b) coefficient $\alpha$ and (c) exponent $\beta$ for the fits are plotted as a function of ScN function and clearly demonstrates that $\varepsilon'$ both increases and gets more dispersive for MBE materials. The sputtered film is not dispersive and has a fixed real permittivity over the entire band.

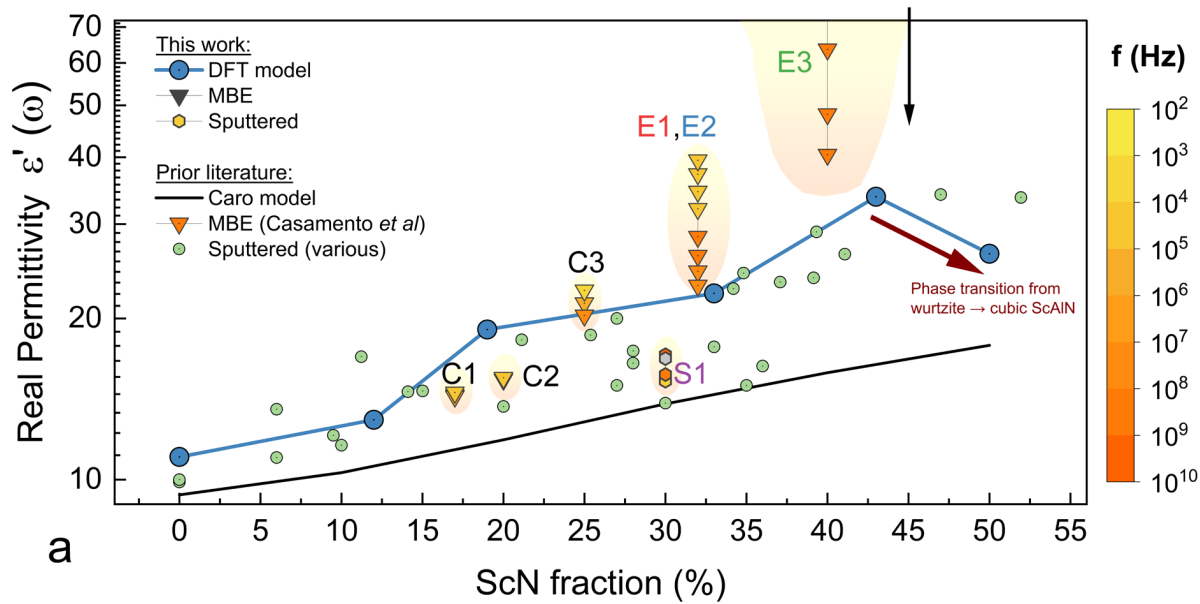
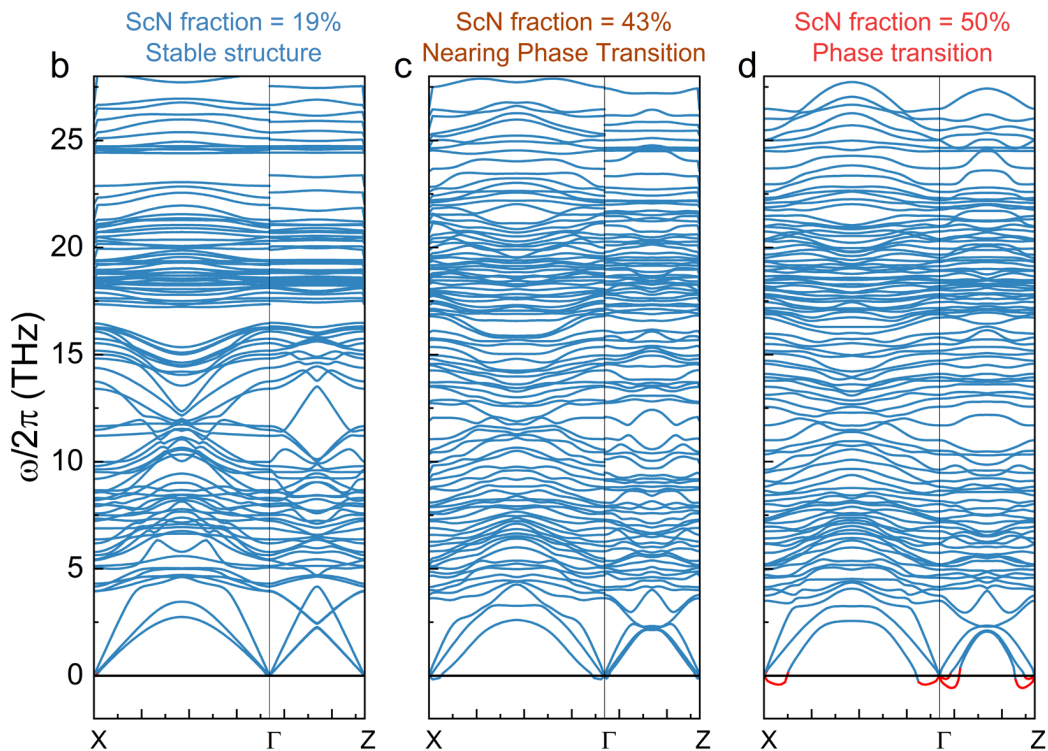

Figure 3: **Density functional theory calculations for permittivity of wurtzite ScAlN.** (a) A comparison of measured data as a function of ScN fraction with *ab initio* numerical modeling using density functional theory (DFT). The DFT calculations predict higher values than the Caro model [5]. Permittivity values reported from sputtered ScAlN in the literature are shown for comparison. (b-d) Phonon band structures of wurtzite ScAlN at various ScN fractions, calculated using DFT. At 43%, the band structure calculations start predicting imaginary modes, indicating a breakdown of the wurtzite structure and a transition to the cubic structure. The clearly visible imaginary modes at 50% ScN fraction underlines the phase instability of ScAlN. Due to reduced symmetry in the wurtzite ScAlN system, the phonon band structures are plotted along the *a*- and *c*- axes.

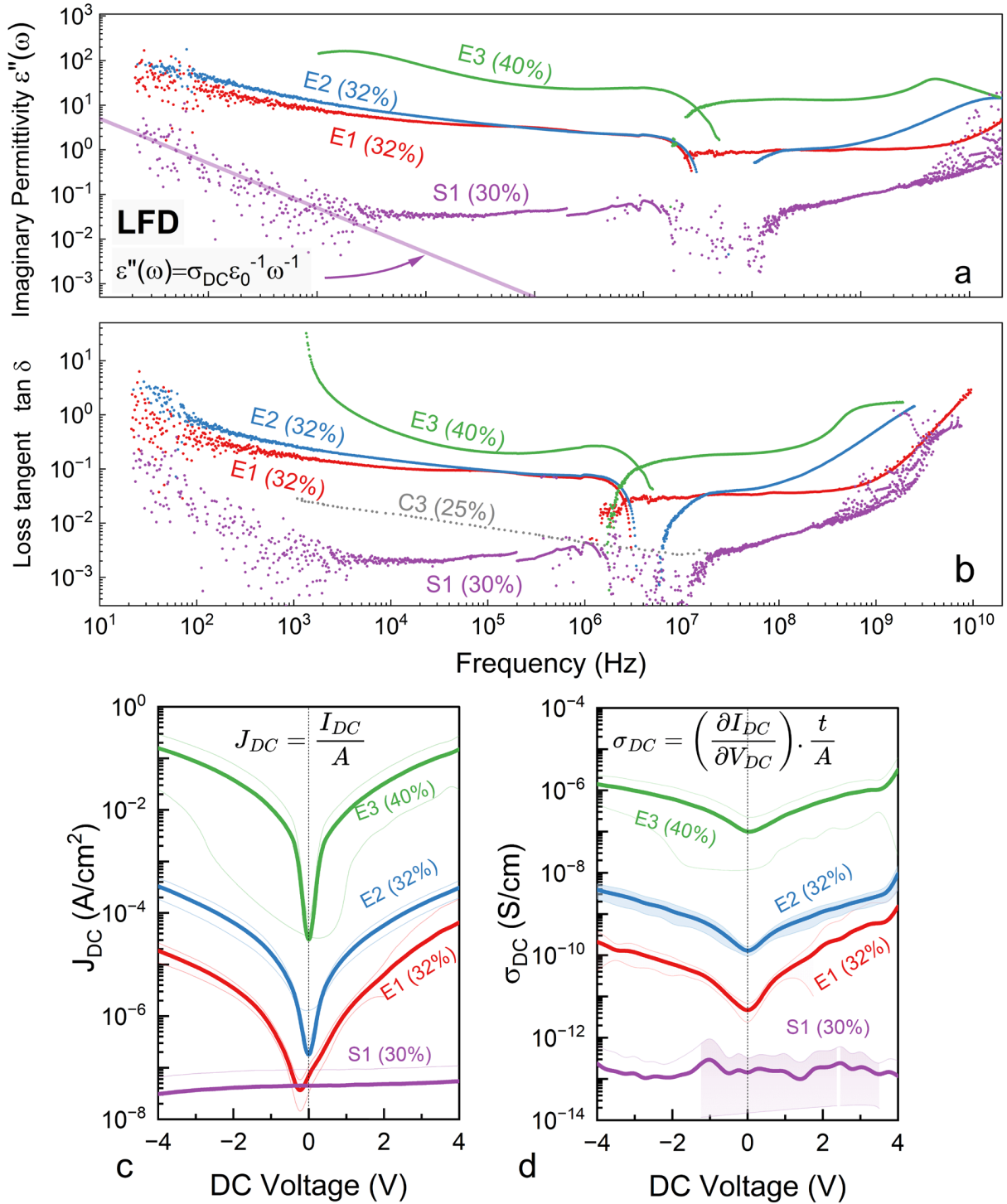

Figure 4: **Imaginary permittivity, loss tangent, and DC leakage in the ScAlN thin films.** Extracted values of (a) imaginary permittivity and (b) loss tangent for the ScAlN films, for MIMCAPs of inner radius $r_1 = 50$ μm, for all samples. (c) DC current density $J_{DC}$ and (d) DC conductivity $\sigma_{DC}$ for all samples, demonstrating clear differences in electrical leakage. Sputtered sample S1 has barely any leakage and can be considered to be practically insulating. The MBE samples all have significantly more leakage, with E3 showing the worst behavior. Solid lines and shaded regions represent average values and standard deviations respectively across all MIMCAPs for a particular sample.

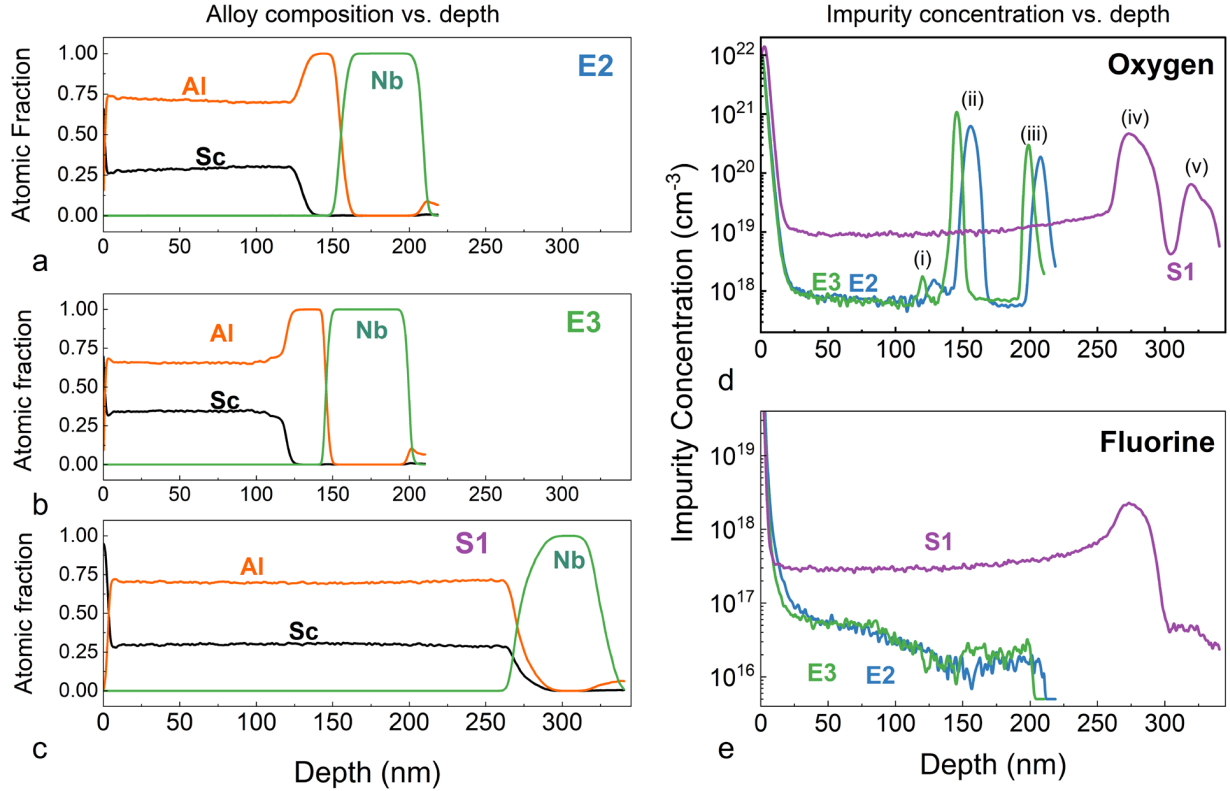

Figure 5: **ScAlN composition and impurity doping profiles.** (a-c) Secondary-ion mass spectroscopy (SIMS) profiles showing alloy composition as a function of depth for E2, E3, and S1 respectively. The nearly flat profiles for Sc and Al in the ScAlN films validate the ability of both deposition techniques to grow high quality ScAlN films. The SIMS data show the AlN nucleation layers in MBE ScAlN and the NbN bottom electrodes in all samples. The ScN fraction is slightly lower than nominal values in all samples but within the measurement error bound for SIMS (± 5%). (d, e) Comparative SIMS profiles for two dopant impurities: oxygen, and fluorine. The oxygen profiles in (d) clearly show that the MBE ScAlN films have ~ 10x lower oxygen content than the sputtered ScAlN, which is expected given the ultra-high vacuum conditions of MBE growth. Thus, just the presence or proportion of impurity oxygen is not sufficient explanation for the higher leakage and dispersive behavior observed in MBE ScAlN films. Interfaces (i) – (iii) are the ScAlN→AlN, AlN→NbN, NbN→SiC interfaces for the MBE ScAlN, whereas (iv) and (v) are the ScAlN→NbN and NbN→SiC interfaces in the sputtered ScAlN. We observe sharp build-up of oxygen at all NbN interfaces, which is expected due to growth interrupts (MBE) or exposure to atmosphere (sputtered), and since Nb surfaces readily oxidize. Fluorine (e) similarly has higher concentration in sputtered ScAlN. We also test for carbon and silicon (Supplementary); both elements are found in nearly the same proportion in all films. All SIMS measurements were carried out at Eurofins EAG Laboratories.

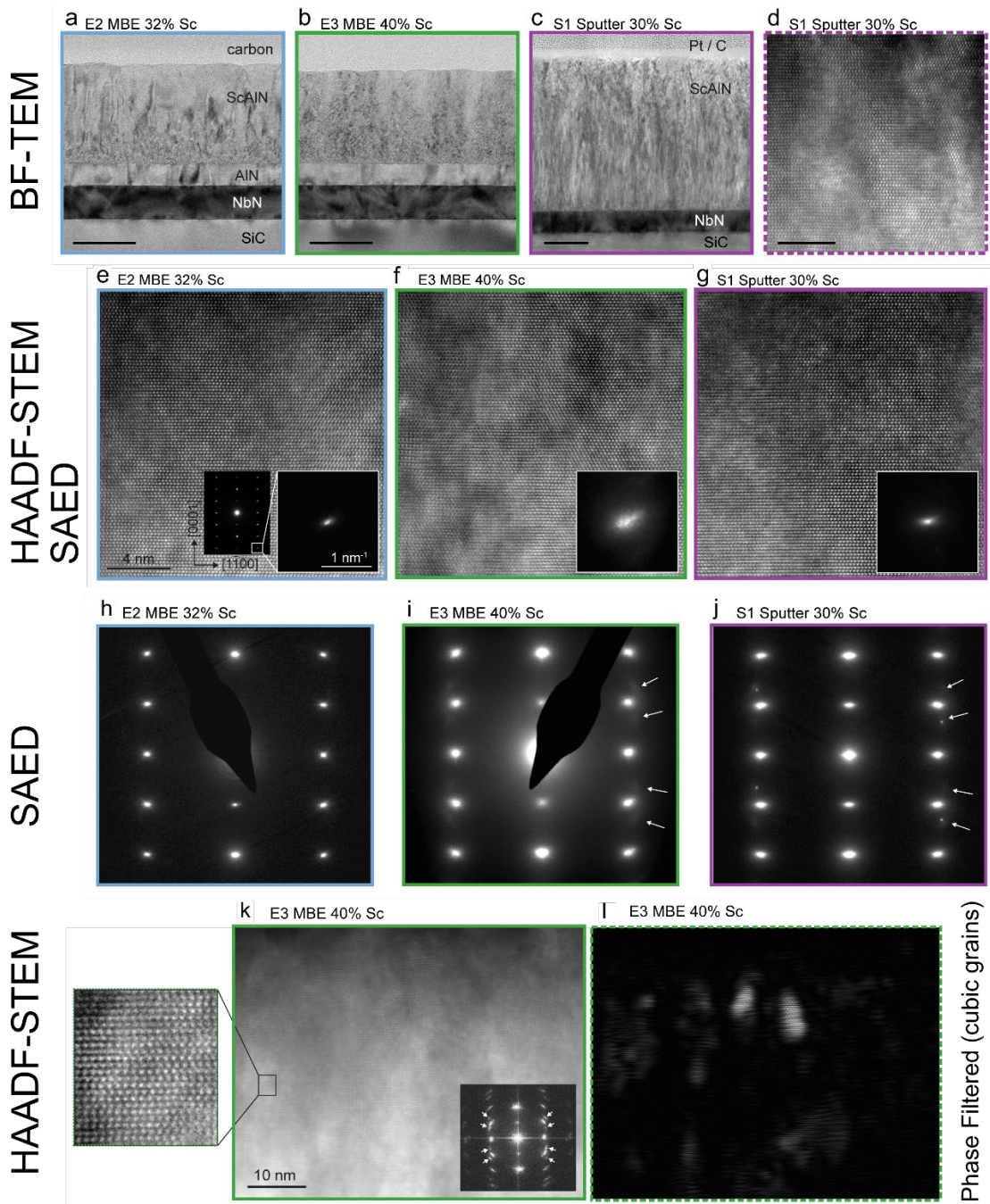

Figure 6: **Film Microstructure.** (a-c) BF-TEM images of the E2, E3, and S1. For each image, the scale bar is 100 nm. (d) HAADF-STEM image of S1, with a scale bar of 5 nm. (e-g) HAADF-STEM images taken down the [11-20] zone axis. All three images have a scale bar of 4 nm (seen in e). Insets show SAED patterns highlighting the $(1\bar{1}0\bar{3})$ reflection. The sharpness of this spot is a proxy for the crystal quality. For each inset, the intensity is normalized to the maximum $(1\bar{1}0\bar{3})$ intensity. (h-j) SAED patterns taken down the [11-20] zone axis. The white arrows mark the cubic reflections. For E3, these reflections are very diffuse and broad, indicating many nanoscale grains. For S1, sharp cubic peaks are also observed, indicating a large cubic grain. (k). Large field-of-view atomic resolution HAADF-STEM image of sample E3. The left inset shows a magnified region of the same image, demonstrating the atomic resolution. The bottom right inset shows a Fourier transform (FT) of the image. White arrows highlight reflections from cubic grains. (l). Phase map of the cubic grains, obtained by masking the spots shown in k and then taking the inverse FT.

# X. Supplementary Information

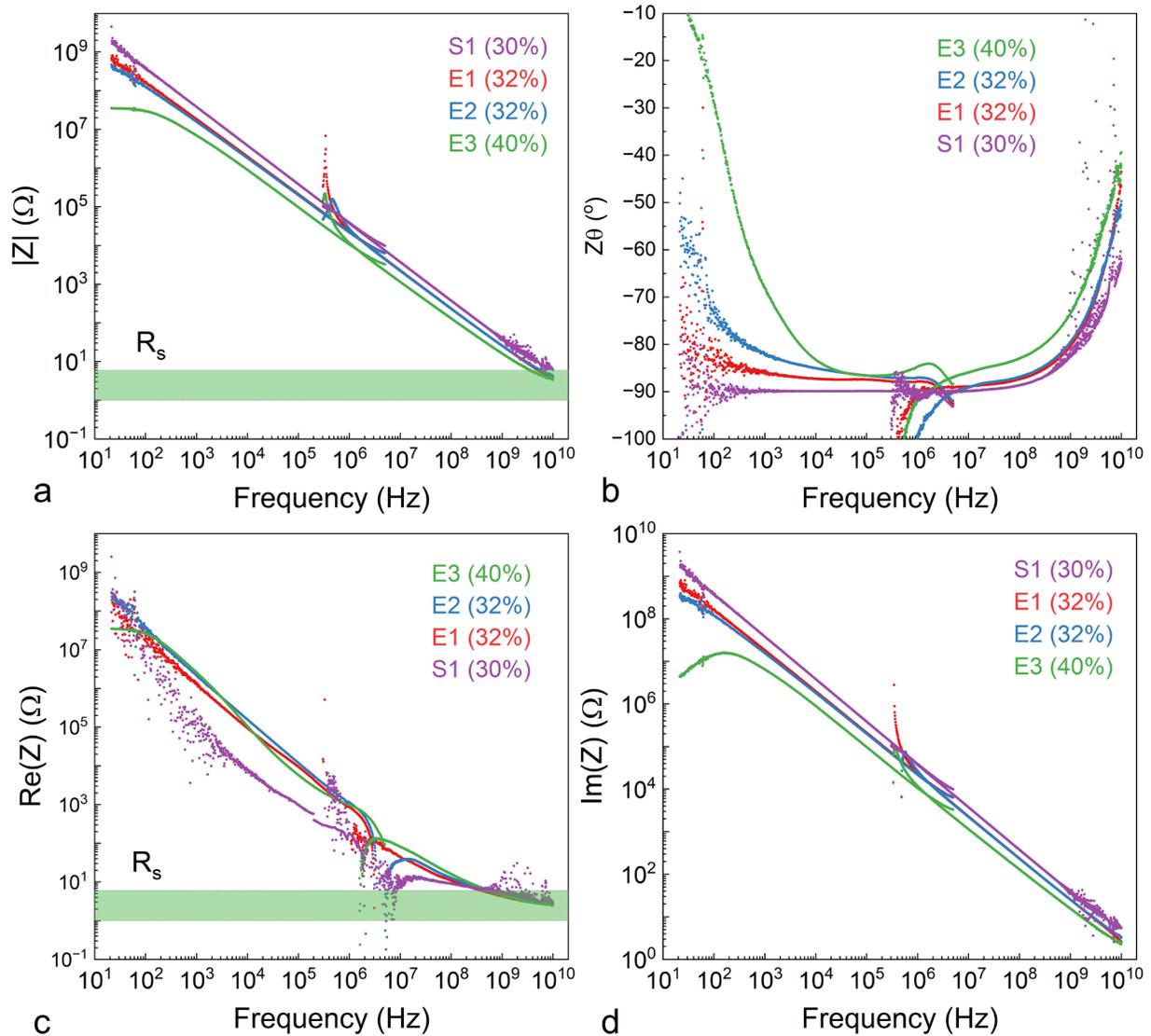

Supp. Fig. 1: As measured impedance data across the entire spectral range for all ScAlN samples, for a representative MIMCAP of inner radius $r_1 = 50$ µm. (a-b) show the magnitude and phase of the impedance, while (c-d) show the same impedance data separated into real and imaginary components. The data are measured using an impedance analyzer at low frequencies and a vector network analyzer at high frequencies. For extraction of the complex permittivity, the only arbitrary free parameter is the contact resistance $R_s$, which assumes values between 1Ω and 6Ω (green band in (a) and (c)). From (a) and (c) it is clear that $R_s$ is negligible compared to the other components of impedance for frequencies below ~ 1GHz

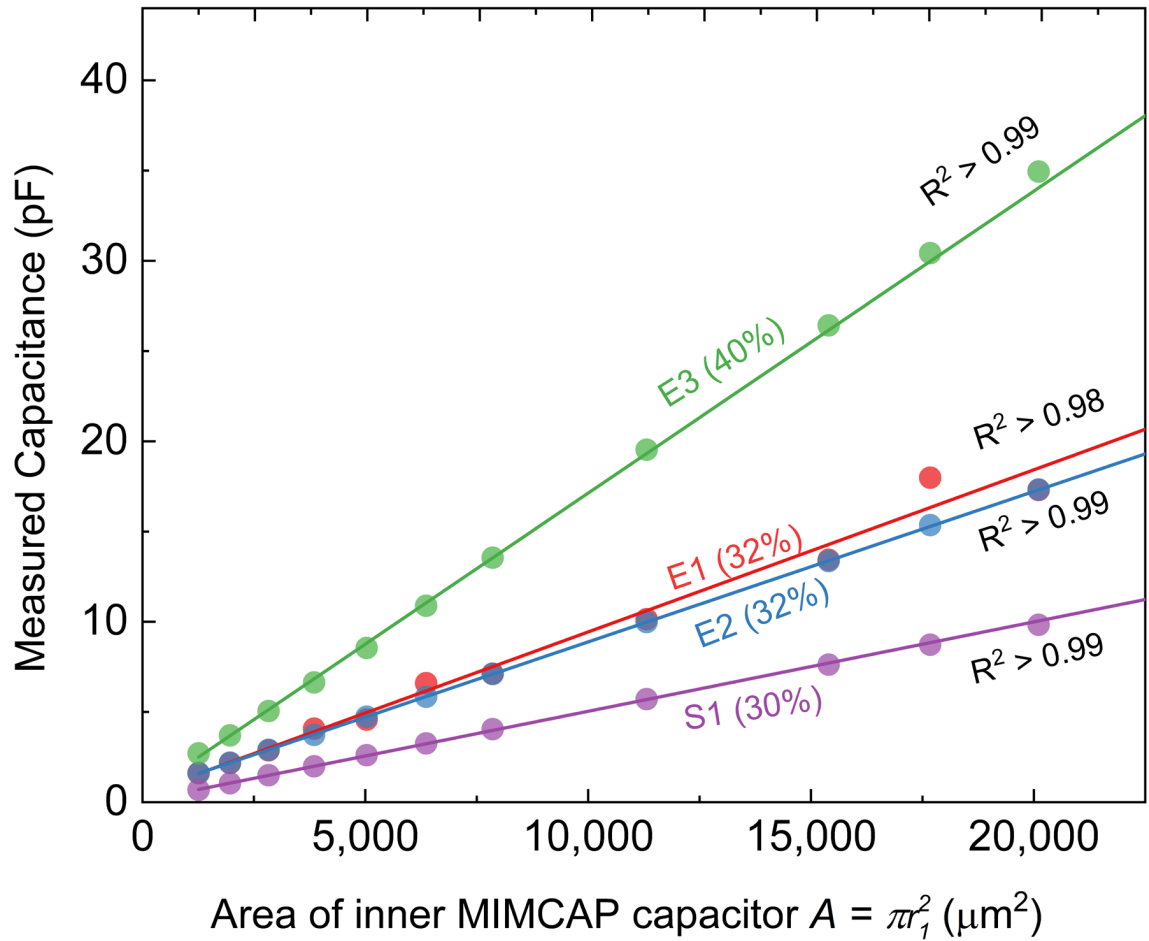

**Supp. Fig. 2:** Measured real capacitance for all ScAlN samples as a function of area of the inner MIMCAP capacitor, given by $A = \pi r_1^2$ and following the expected and well-known relation $C = A\varepsilon_0\varepsilon/t$ for a parallel plate capacitor. Data are averaged across all frequencies for a given MIMCAP device.

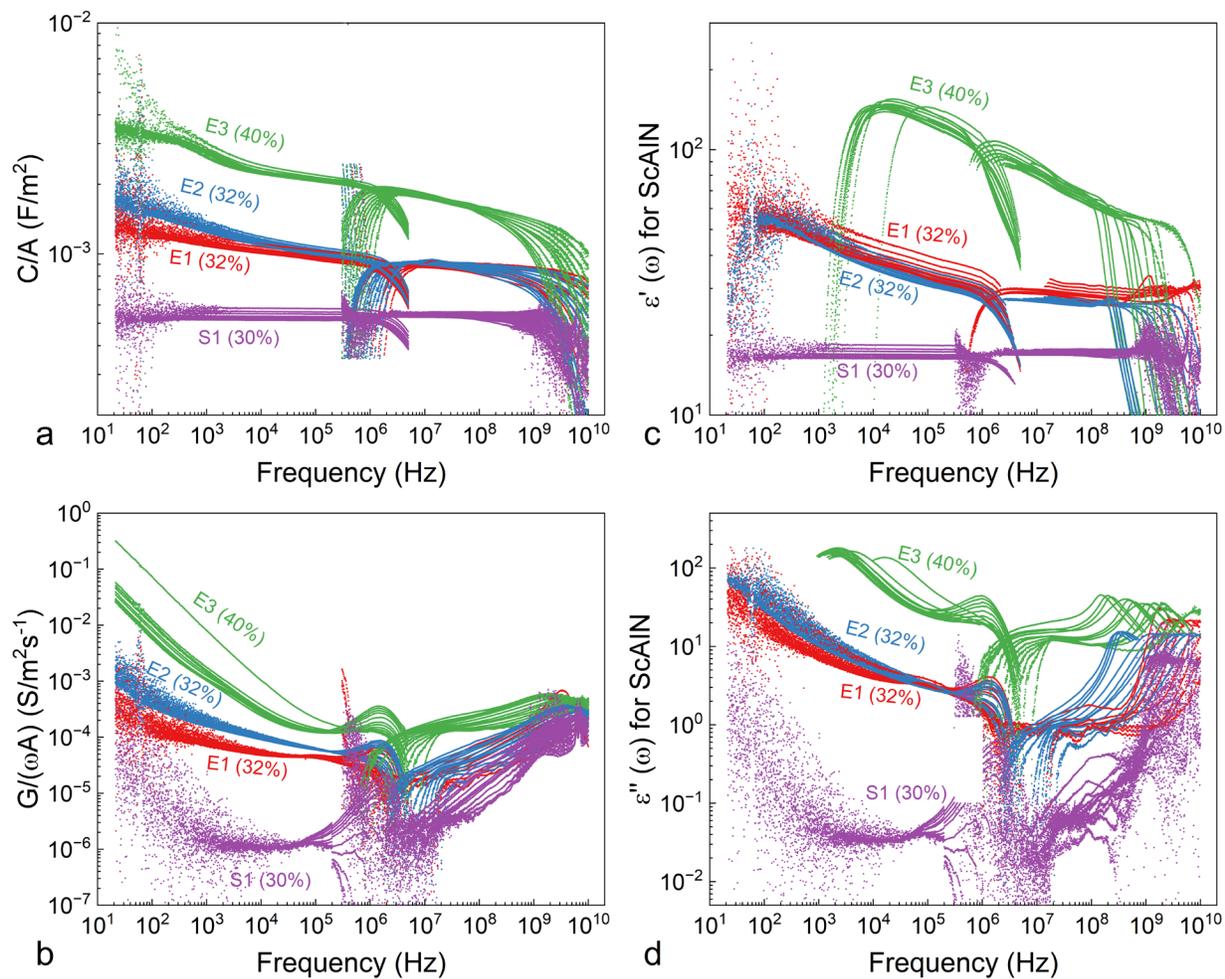

**Supp. Fig. 3:** Data for all MIMCAPs used in this work, with inner electrode radius $r_1$ ranging from 15 μm to 85 μm, for all MBE samples (E1 – E3) and sputtered sample (S1). (a-b) shows the measured and normalized data for capacitance and conductance $(C/A)$ and $(G/\omega A)$ respectively, while (c-d) shows the extracted real permittivity $\varepsilon'(\omega)$ and imaginary permittivity respectively $\varepsilon''(\omega)$.

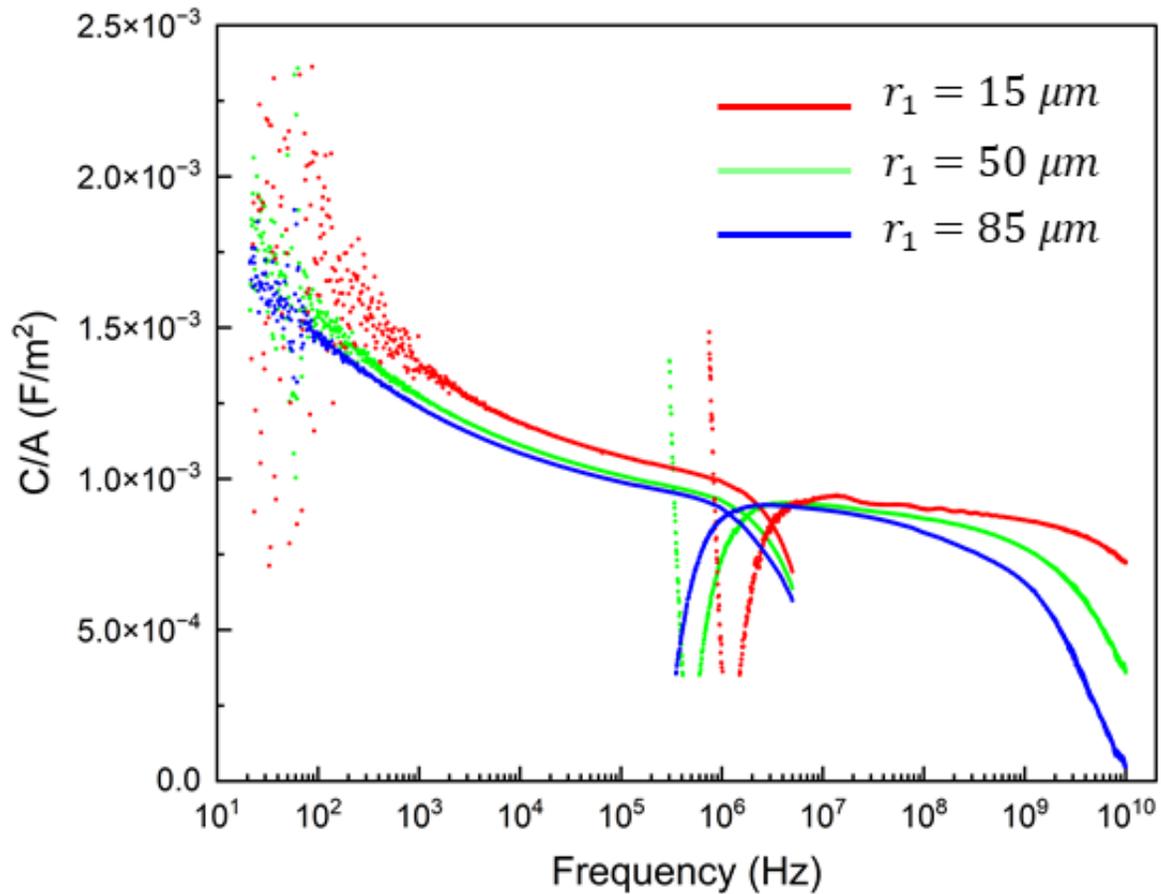

**Supp. Fig. 4:** Measured and normalized capacitance ($C/A$) for sample E2, for MIMCAPs with nominal inner radii $r_1$ = 15 µm, $r_1$ = 50 µm, and $r_1$ = 85 µm. The figure clearly shows that the smaller MIMCAPs have a larger cutoff frequency, and thus a larger measurement range. On the other hand, smaller MIMCAPs are noisier at low frequencies while the larger MIMCAP have more reliable data at low frequencies. This experiment highlights the importance of measuring and extracting data from an array of MIMCAPs of varying size. There is a small observed offset in $C/A$ that is a result of lithographic errors; the radii of the fabricated MIMCAPs are slightly smaller than the nominal values, and the effect is starker for the smaller MIMCAPs.

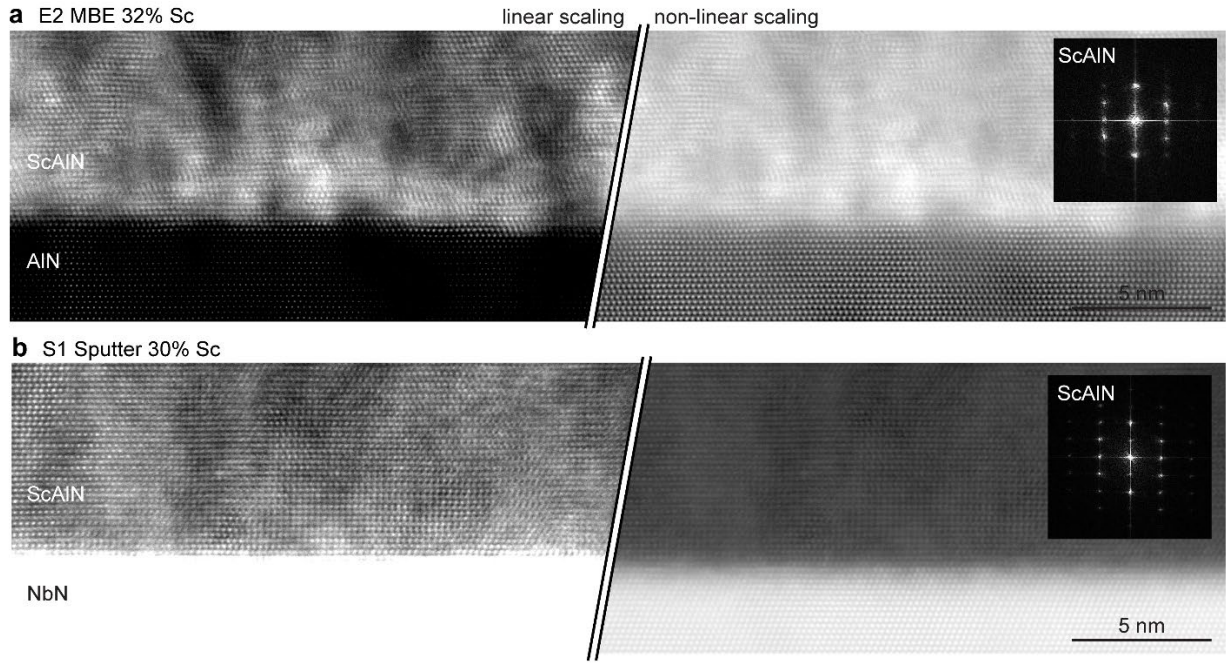

**Supp. Fig. 5:a-b** HAADF-STEM images for the ScAlN / substrate interface for sample E2 and S1, respectively. (a) shows the ScAlN / AlN interfaces for E2 and (b) shows the ScAlN / NbN interface for S1. In the left-side of both images, the contrast limits are set to evaluate the ScAlN atomic structure. On the right side of the image, the contrast limits are set to visualize both the ScAlN and the substrate, using non-linear scaling to capture the large difference in image intensity. The insets show the FTs of ScAlN layer directly adjacent to the interface. While E2 shows the highest crystal quality within the film interior, the sputtered film S1 has the sharpest interface. For E2, the ScAlN crystal quality at the interface is clearly much lower than the crystal quality 100 nm into the film. A Fourier transform (FT) of the ScAlN shows the expected [11-20] wurtzite structure; however, the spots are quite broad, and only low-order spots are visible. The increased interfacial disorder in E2 is also observed in the low magnification BF-TEM.

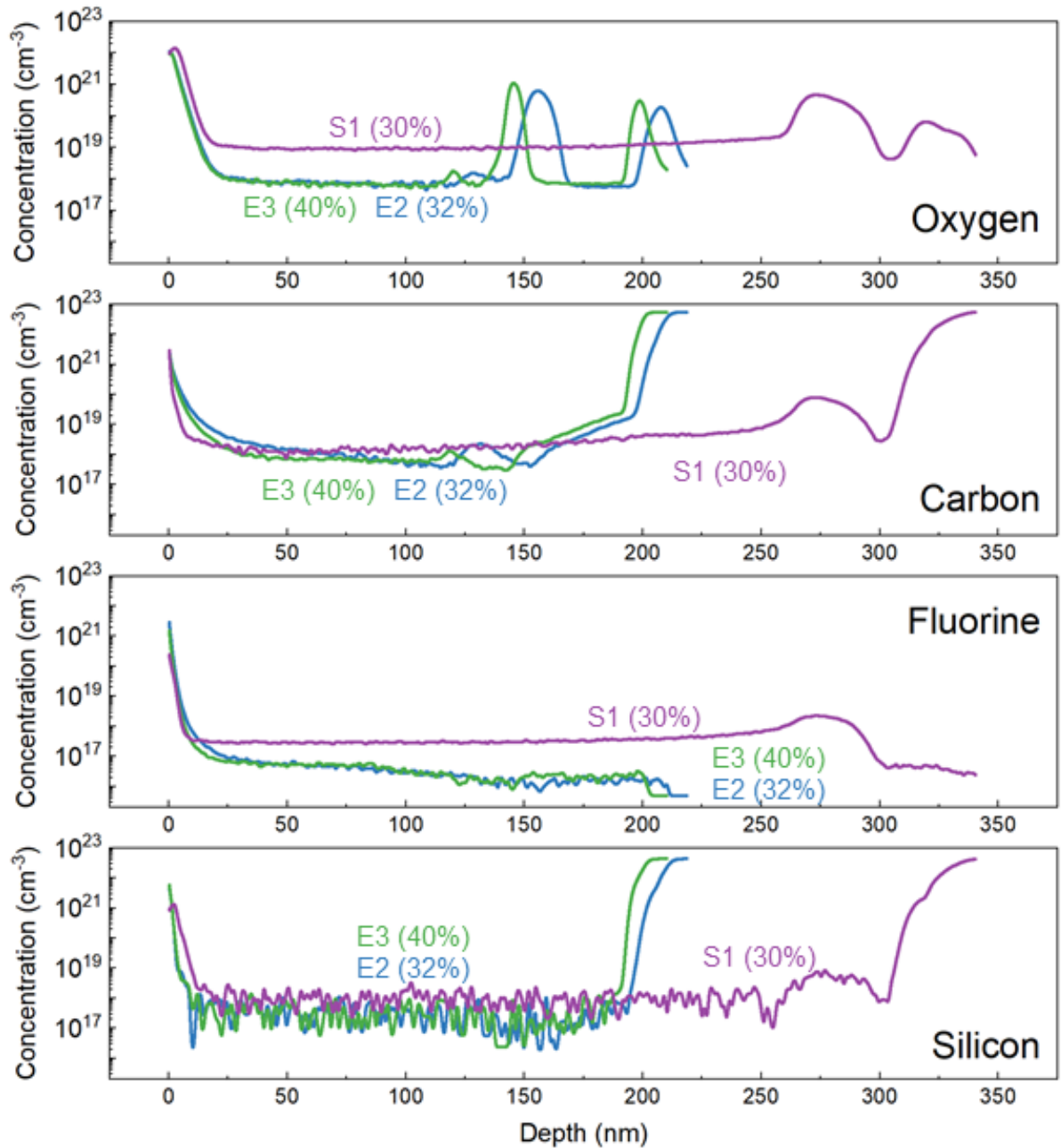

**Supp. Fig. 6:** Impurity doping profiles for oxygen, carbon, fluorine, and silicon in samples E2, E3, and S1. Both oxygen and fluorine levels are higher in S1 than E2 or E3. There is no significant difference in carbon or silicon profiles in the ScAlN samples.